\documentclass[twocolumn]{aastex631}

\newcommand{\rl}{$R_{\rm BLR} - L$}
\newcommand{\msigma}{$M_{\rm BH}-\sigma_{\star}$}

\newcommand{\mbh}{$M_{\rm BH}$}
\newcommand{\sersic}{S\'{e}rsic}

\newcommand{\msun}{$M_{\odot}$}

\shorttitle{Reverberation Mapping of NGC\,3227}
\shortauthors{Bentz et al.}

\begin{document}

\title{Velocity-Resolved Reverberation Mapping of NGC\,3227}


\author[0000-0002-2816-5398]{Misty C.\ Bentz}
\affiliation{Department of Physics and Astronomy,
		 Georgia State University,
		 Atlanta, GA 30303, USA}
\email{bentz@astro.gsu.edu}

\author[0009-0000-9670-2194]{Madison Markham}
\affiliation{Department of Physics and Astronomy, 
        Colgate University, 
        13 Oak Drive, Hamilton, NY 13346, USA}

\author[0000-0001-5055-507X]{Sara Rosborough}
\affiliation{School of Physics and Astronomy, 
    and Laboratory for Multiwavelength Astrophysics, 
    Rochester Institute of Technology, 
    Rochester, NY 14623, USA}

\author[0000-0003-0017-349X]{Christopher A.\ Onken}
\affiliation{Research School of Astronomy and Astrophysics, 
        Australian National University, 
        Canberra, ACT 2611, Australia}
        
\author[0000-0001-6279-0552]{Rachel Street}
\affiliation{LCOGT, 6740 Cortona Drive, Suite 102, 
        Goleta, CA 93117, USA}

\author[0000-0002-6257-2341]{Monica Valluri}
\affiliation{Department of Astronomy,
         University of Michigan,
         Ann Arbor, MI, 48109, USA}

\author[0000-0002-8460-0390]{Tommaso Treu}
\altaffiliation{Packard Fellow}
\affiliation{Department of Physics and Astronomy, 
    University of California, 
    Los Angeles, CA 90095, USA}

\begin{abstract}

We describe the results of a new reverberation mapping program focused on the nearby Seyfert galaxy NGC\,3227.  Photometric and spectroscopic monitoring were carried out from 2022 December to 2023 June with the Las Cumbres Observatory network of telescopes.  We detected time delays in several optical broad emission lines, with H$\beta$ having the longest delay at $\tau_{\rm cent}=4.0^{+0.9}_{-0.9}$\,days and \ion{He}{2} having the shortest delay with $\tau_{\rm cent}=0.9^{+1.1}_{-0.8}$\,days. We also detect velocity-resolved behavior of the H$\beta$ emission line, with different line-of-sight velocities corresponding to different observed time delays. Combining the integrated H$\beta$ time delay with the width of the variable component of the emission line and a standard scale factor suggests a black hole mass of $M_{\rm BH}=1.1^{+0.2}_{-0.3} \times 10^7$\,\msun. 
Modeling of the full velocity-resolved response of the H$\beta$ emission line with the phenomenological code {\tt CARAMEL} finds a similar mass of $M_{\rm BH}=1.2^{+1.5}_{-0.7} \times 10^7$\,\msun, and suggests that the H$\beta$-emitting broad line region (BLR) may be represented by a biconical or flared disk structure that we are viewing at an inclination angle of $\theta_i\approx33\degr$ and with gas motions that are dominated by rotation.  The new photoionization-based BLR modeling tool {\tt BELMAC} finds general agreement with the observations when assuming the best-fit {\tt CARAMEL} results, however {\tt BELMAC} prefers a thick disk geometry and kinematics that are equally comprised of rotation and inflow. Both codes infer a radially extended and flattened BLR that is not outflowing.
\end{abstract}


\section{Introduction} 

The prevalence of supermassive black holes with masses of $10^6 \lesssim M_{\rm BH} / M_{\odot} \lesssim 10^{10}$ in the centers of massive galaxies was one of the surprise discoveries enabled by the Hubble Space Telescope (e.g., \citealt{ford94,harms94,kormendy95,vandermarel97}).  More surprising still was the finding that the masses of these black holes scale with observable properties of their host galaxies (e.g., \citealt{ferrarese00,gebhardt00}).  After three decades of study, we now understand that these black holes generally go through at least one period of active accretion during their lifetimes, and that the accretion process releases huge amounts of energy across the electromagnetic spectrum, energy that is deposited into the gas within the host galaxy and circumgalactic environment and thus seems to play a role in modulating the growth of the galaxy and of the black hole itself (see the reviews by, e.g., \citealt{fabian12,heckman14,fan23}).

The mass of a black hole is one of its fundamental properties, and it sets the limit for the amount of feedback power that may be released during an accretion event.  Thus it is an important measurement to constrain as we attempt to further understand the detailed  physics involved in black hole feeding and feedback.  Sagittarius A$^{*}$ is the nearest supermassive black hole, and its mass has been precisely determined through monitoring the orbits of individual stars via their proper motions over decade timescales \citep{ghez00,genzel00,ghez08}. Unfortunately, all other galaxies are so distant that we cannot achieve the necessary spatial resolution in their nuclei to apply the same technique.  

A variety of techniques have been developed to determine black hole masses in other galaxies.  These include dynamical modeling of the bulk motions of stars or gas (see the review by \citealt{kormendy13}) and kinematic studies of masing gas clouds (e.g., \citealt{miyoshi95}).  High spatial resolution is a critical component of these techniques, which generally limits their application to galaxies with $D\lesssim100$\,Mpc.

Black hole masses at cosmological distances are most often determined through reverberation mapping (RM; see the recent review by \citealt{cackett21}), a technique that makes use of the photoionized gas deep in the potential well of an actively accreting black hole.  RM utilizes spectrophotometric monitoring with high temporal sampling to track changes in the continuum emission (expected to arise from the accretion disk) and the ``echoes'' of those changes in the broad emission-line fluxes.  The time delay between the two gives the typical light travel time across the broad line region, and thus a physical size.  With reverberation mapping, spatial resolution is replaced by temporal resolution, and black holes at significant distances become available for study.

One effect of having these varied techniques and their disparate limitations is that we have two black hole mass scales currently in use: one determined from nearby, mostly early-type inactive galaxies based on stellar and gas dynamical modeling, and another based on reverberation mapping of more distant active galaxies that tend to be mostly later morphological types.  Because broad-lined active galaxies with $D\lesssim 100$\,Mpc are rare, there are very few galaxies that meet the requirements for multiple techniques that would allow us to check that these two mass scales are in agreement.

NGC\,3227 is a nearby ($z=0.00377$) spiral galaxy that is interacting with its elliptical companion, NGC\,3226.  Both galaxies in the pair show evidence for nuclear activity, a fact that was first remarked upon by \citet{curtis1918} who noted that the pair of galaxies contained ``stellar nuclei''.   NGC\,3226 is classified as a LINER \citep{veroncetty06}, while NGC\,3227 was one of the original sample of 12 \citet{seyfert43} galaxies.  The broad hydrogen lines in the nucleus of NGC\,3227, compared to the widths of the nebular lines, was noted by \citet{dibai68} and \citet{rubin68}, though other early studies have sometimes classified it as a Type II Seyfert (e.g., \citealt{khachikian74}).  

Similarly, early studies were divided regarding the potential detection of nuclear variability in NGC\,3227 (e.g., \citealt{netzer74,liutyi77}).  While \citet{tennant83} detected  X-ray variability on timescales of $\sim 1$\,day in NGC\,3227, \citet{salamanca94} were the first to systematically monitor the AGN with optical spectroscopy on short timescales and demonstrate clear variability in the continuum and broad emission lines. With only 26 measurements over the course of 6 months, however, their light curves were severely undersampled when it came to trying to constrain any time delays between them.  A similar campaign with comparable time sampling described by \citet{winge95} was also plagued by the same difficulties.  A re-analysis of both datasets by \citet{peterson04} using a set of best practices that had been developed for reverberation mapping studies concluded that any potential time delays were formally consistent with zero because of the coarse temporal sampling.

\citet{denney10} thus undertook a new monitoring program with the intent of determining the first accurate and precise BLR time delay measurement for NGC\,3227, finding that the H$\beta$ broad line emission lagged the continuum variations by $3.8\pm0.8$\,days and showed evidence for different time delays as a function of velocity across the emission line profile. Further campaigns by \citet{derosa18} and \citet{brotherton20} sought to improve on these results by more fully exploring the velocity-resolved time delays to determine the BLR geometry and kinematics, in the hopes of deriving a more direct constraint on the black hole mass.  Preliminary modeling of the \citet{derosa18} data sets suggest that the low amplitude of variability in the light curves may hinder any firm conclusions from being drawn \citep{robinson22}, and while \citet{brotherton20} provide a velocity-delay map for H$\beta$ based on their monitoring data, the lack of details in the map  limited any potential for interpretation.  

We thus undertook a new monitoring program focused on NGC\,3227, which we present in this work.  Strong variability coupled with high temporal sampling allowed us to measure broad emission-line time delays and to further explore the velocity-resolved response of H$\beta$.  These promising  characteristics also enabled a more thorough exploration of the constraints that may be placed on the structure and kinematics of the BLR in this nearby AGN, based on two independent modeling codes that have been developed specifically for use with reverberation mapping data.
 
\section{Observations}

NGC\,3227 is located in the direction of the constellation Leo at $\alpha=10$h23m30.6s and $\delta=+19\degr51\arcmin54\arcsec$ with a redshift of $z=0.00377$.  A surface brightness fluctuations (SBF) distance to the companion galaxy NGC\,3226 puts the pair at a distance of $D=23.7\pm2.6$\,Mpc (\citealt{tonry01} with the adjustments of \citealt{blakeslee01}).

\subsection{Imaging}

Broad-band photometric monitoring of NGC\,3227 began on 2022 December 01 and continued through 2023 June 04 (UT dates here and throughout) under programs LCO-2022B-003 and LCO-2023A-004.  Observations were collected on the 1m telescope network of Las Cumbres Observatory (LCO; \citealt{brown13}) using the Sinistro imagers equipped with Johnson $V$ filters.  Exposures were scheduled to occur every $\sim 8$\,hr with a typical exposure time of 60\,s.  Each image covers a field of view of $26\farcm5 \times 26\farcm5$ with a pixel scale of 0\farcs389.

A total of 376 images were collected from 5 different observatory sites over the course of the monitoring program:  93 at Siding Spring Observatory (SSO), 85 at South African Astronomical Observatory (SAAO), 84 at McDonald Observatory, 65 at Cerro Tololo Interamerican Observatory (CTIO), and 49 at Teide Observatory. Observations were automatically reduced by the LCO pipeline \citep{mccully18}, with typical CCD reductions using biases, darks, and flats, before being downloaded from the LCO archive\footnote{https://archive.lco.global}.

Rather than rely on aperture photometry, which may include large amounts of host-galaxy starlight that varies with the seeing and also dampens the true brightness variations of the central AGN, we instead used the image subtraction algorithms of \citet{alard98,alard00}.  In general, image subtraction works by scaling individual images to match a reference image and then subtracting the two, removing all sources of constant flux and leaving behind only the variable flux.  Image subtraction methods are commonly applied in studies of extragalactic transients and other time domain phenomena, including microlensing \citep{udalski08,udalski15}, supernovae  \citep{riess01,miknaitis07,melinder08}, tidal disruption events \citep{holoien16,brown18}, as well as AGN variability \citep{fausnaugh16,grier17,bentz21a}.  

\begin{figure}
    \epsscale{1.1}
    \plotone{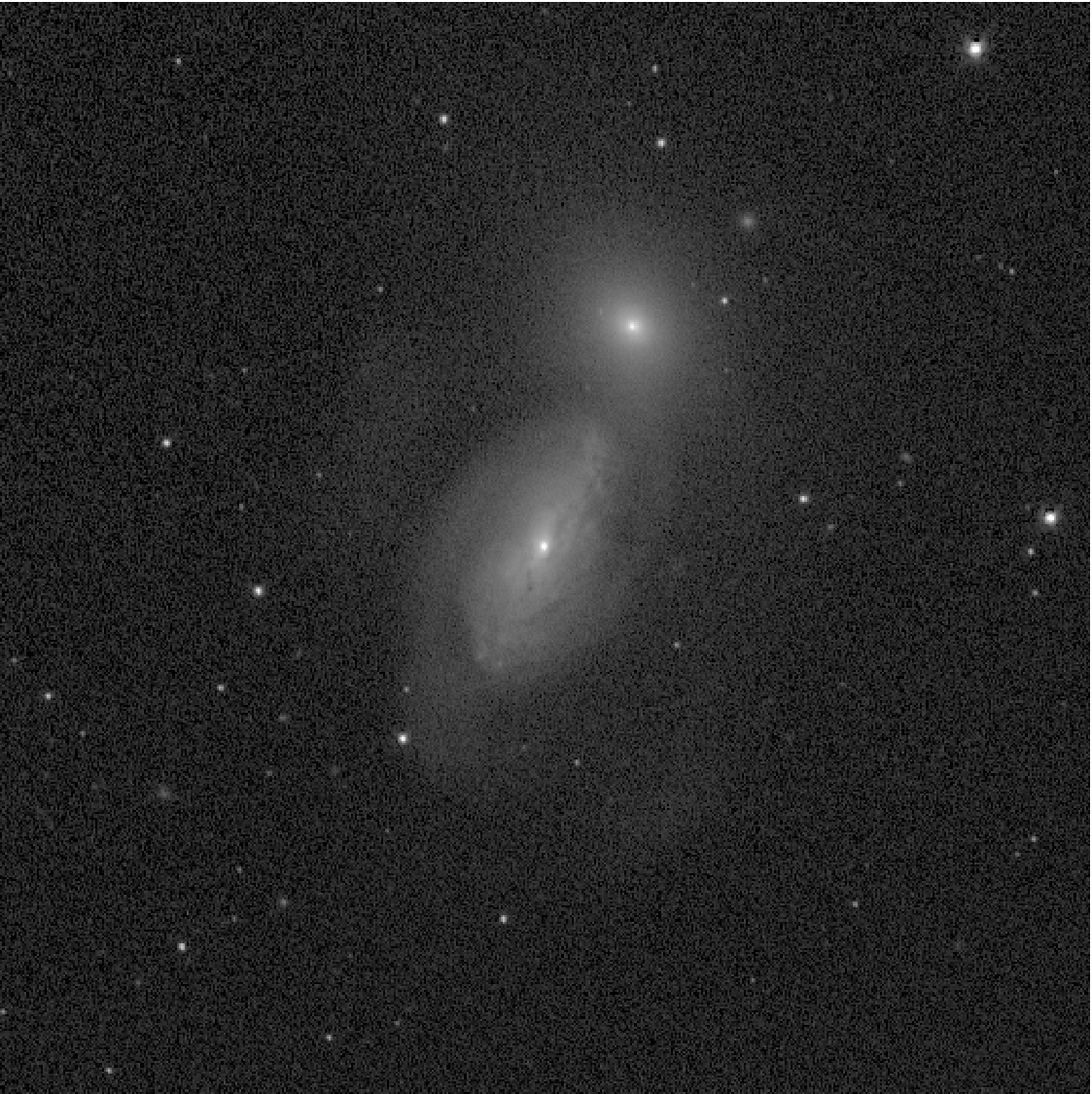}
    \caption{
    The inner $10\arcmin\times10\arcmin$ region of the reference image, centered on the nucleus of NGC\,3227 and oriented with North up and East to the left.
    }
    \label{fig:refimg}
\end{figure}

We began by registering all of the images to a common grid using the algorithm of \citet{siverd12}. We then selected the highest quality images (lowest background and best seeing) from the subset that were obtained at SSO, and stacked them to build a reference image (see \ref{fig:refimg}).  The reference image was convolved to match, and then subtracted from, each of the 376 individual images.  Finally, aperture photometry with a radius of 9 pixels was performed at the location of the AGN in all of the subtracted images, giving us a light curve of the variable AGN counts.

While the LCO network employs identical telescopes and detectors across sites, we have previously found slight offsets between the measurements obtained at different LCO locations \citep{bentz21a,bentz23}. Accordingly, we compared the light curves obtained at each location to the light curve obtained at SSO and again found that, when working in counts, small adjustments were necessary to bring them all into agreement.  For a pair of light curves, we identified the set of measurements that were taken close together in time (usually within $0.5-1$\,days) and determined the slope and intercept of the relationship between the measurements from one observatory versus the other, taking into account the errors in both measurements.  Good agreement would be indicated by a slope of 1.0 and intercept of 0.0, with some scatter given the small and random variations of the AGN on very short timescales.  In all comparisons, the slope and intercept were found to deviate somewhat from these expected values, demonstrating a difference in the counts that are registered on the detectors at different observatory locations, likely due to factors like slight variations in the overall sensitivities of the CCDs or differences in atmospheric transparency.  We applied a small additive and multiplicative correction factor to each light curve to bring them into agreement with the SSO light curve, which was the basis for the reference image.

We then calibrated the light curves using the AGN flux in the reference image.  To separate the AGN from its bright and spatially extended host galaxy, we employed the two-dimensional surface brightness modeling package {\tt Galfit} \citep{peng02,peng10}.  We first constructed a PSF model for the reference image by fitting multiple Gaussians to a small cutout centered on an isolated field star, with the background sky brightness modeled as a gradient in $x$ and $y$.  Four Gaussians were found to be sufficient to capture the detailed shape of the PSF.  We then modeled the reference image using this PSF to simulate the AGN and several field stars, a gradient in $x$ and $y$ for the background sky component, and multiple \sersic\ profiles for the galaxies NGC\,3227 and NGC\,3226.  Our choice of starting parameters for the \sersic\ profiles was guided by previous surface brightness models of NGC\,3227 using high-resolution Hubble Space Telescope images \citep{bentz09b,bentz18}.  Once we had a good model that allowed us to cleanly separate the AGN flux from the surrounding host galaxy, we determined the final calibration of the reference frame by setting the zeropoint such that the recovered $V-$band magnitudes for several field stars matched the values tabulated by the AAVSO Photometric All Sky Survey catalog \citep{henden14} DR10.  The calibrated $V-$band magnitude for the AGN in the reference image was then combined with the variable flux measured from image subtraction to determine the calibrated $V-$band light curve. 

We next examined the residual counts around several non-varying field stars in the subtracted images and compared these measurements to the uncertainties reported on the aperture photometry at the location of the AGN.  Previous work by \citet{zebrun01,hartman05} and others has demonstrated that the uncertainties in image subtraction measurements are often underestimated, and we found that to hold true in the case of our analysis of NGC\,3227.  We followed the basic procedure outlined by \citet{hartman04} to determine that a multiplicative factor of $\sim 9$ needed to be applied to provide more realistic uncertainties.  

\begin{figure}
    \epsscale{1.15}
    \plotone{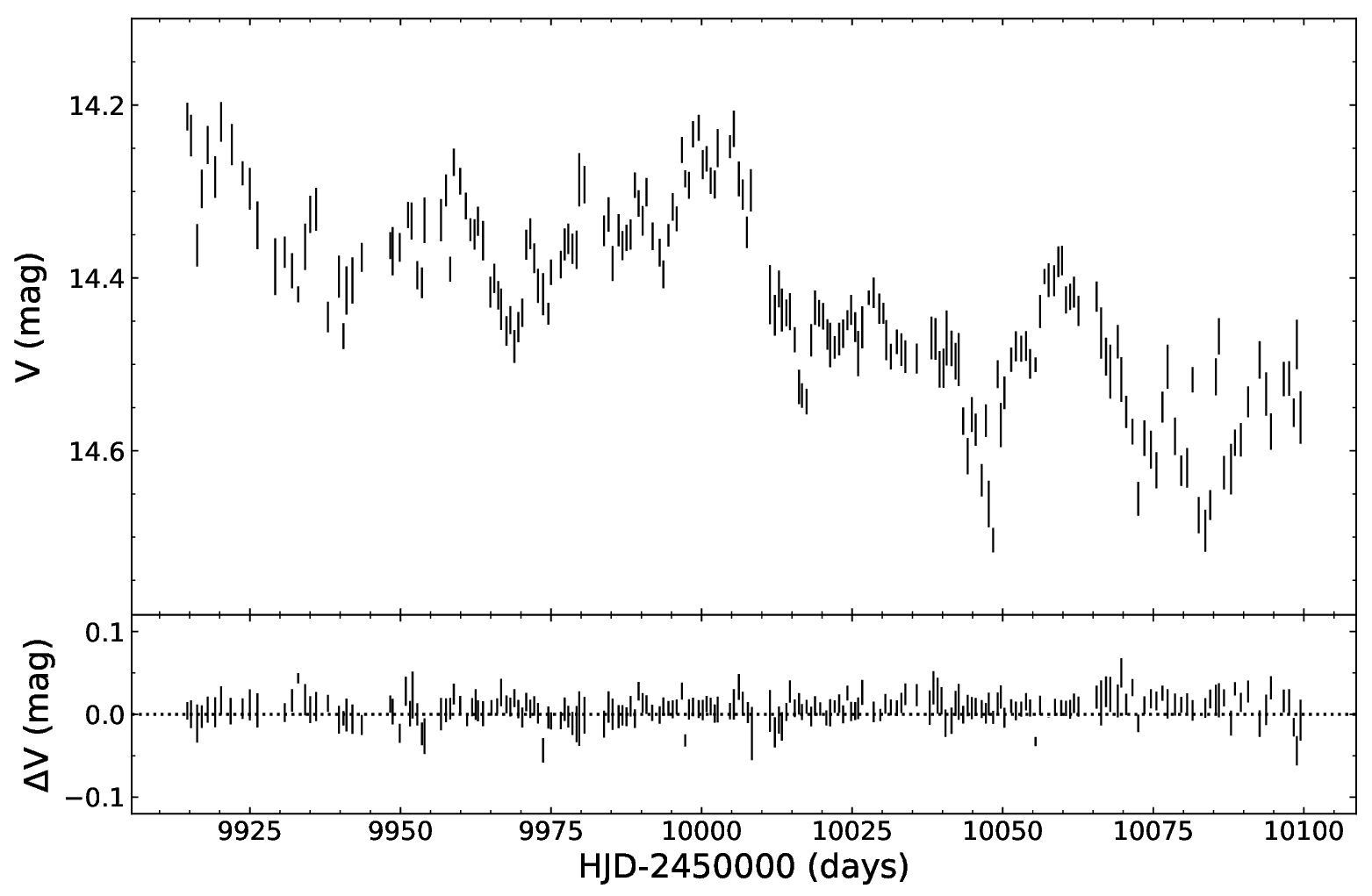}
    \caption{
    The final calibrated and binned $V-$band light curve for NGC\,3227.  For comparison with the AGN variations and uncertainties, the bottom panel shows the scatter in residuals for a non-varying field star of similar brightness.
    }
    \label{fig:vbandlc}
\end{figure}

Finally, we examined the effectiveness of binning the light curve for decreasing some of the noise that may result from images that were obtained under marginal observing conditions.  We found that a bin size of 0.5\,days was a good compromise between retaining much of the temporal sampling while also decreasing some of the noise. In Figure~\ref{fig:vbandlc}, we show the final calibrated and binned $V-$band light curve for the central AGN in NGC\,3227 during this monitoring campaign.  The bottom panel shows the scatter in the residuals for a non-varying field star of similar magnitude for comparison with the AGN variability.  The light curve is tabulated in Table~\ref{tab:lc}, with flux densities based on the absolute spectrophotometric calibration of Vega provided by \citet{colina96}.

\subsection{Spectroscopy}

Spectroscopic monitoring was carried out on the LCO 2m telescope network between UT dates 2023 February 02 and 2023 June 01 under programs LCO-2023A-004 and ANU-2023A-002.  The FLOYDS cross-dispersed spectrographs provided wavelength coverage of $5400$\,\AA$- 1.0\,\mu$m with a dispersion of 3.51\,\AA/pix in the first order, and wavelength coverage of $3200-5700$\,\AA\ in the second order with a dispersion of 1.74\,\AA/pix.  Spectra were obtained through a 6\arcsec\ slit oriented N-S along a position angle of 0\degr.  Observations were scheduled every $\sim24$\,hr, and a typical visit consisted of a 900\,s science exposure together with an arc lamp and a flat field  taken through the same slit.

A total of 51 spectra were acquired over the course of the program, with 42 acquired by Faulkes Telescope South and 9 by Faulkes Telescope North.  Observations were downloaded from the LCO archive after being processed by the LCO pipeline, which splits the 2D spectra into separate orders, rectifies each order, and applies typical CCD reductions including biases, darks, and flats.  The pipeline also provides initial wavelength and flux calibrations based on stored wavelength and sensitivity functions.  Focusing on the 2nd order spectra, which have a higher spectral resolution and also include the [\ion{O}{3}] $\lambda \lambda 4959,5007$\,\AA\ doublet (see discussion below), we began with the wavelength- and flux-calibrated 2D spectra and carried out cosmic ray cleaning, spectral extraction through a 10\,pix extraction width, and improved wavelength calibration using the arcs taken at the same pointing as each spectrum.  We then adopted a common dispersion of 1.6\,\AA\ for all spectra and cropped them to have the same starting and ending wavelengths.

Next, we applied the spectral scaling method of \citet{vangroningen92} to account for slight variations in the spectra caused by imprecise flux calibration, varying weather and seeing conditions, and slight differences in telescope pointing.  The algorithm applies small shifts in wavelength and flux and a small amount of smoothing to minimize the differences in each spectrum relative to a reference spectrum, where the reference spectrum is generally created from a subset of the best-quality spectra.  We applied the algorithm focusing on the spectral region around the [\ion{O}{3}] $\lambda \lambda 4959,5007$\,\AA\ emission lines, as they are known to not vary on the short timescales that are covered by a single observing season \citep{peterson13}.  This allows the [\ion{O}{3}] doublet to be used as a set of internal comparison sources and also explains why most reverberation mapping experiments have focused on the H$\beta$ emission line: even though H$\alpha$ is brighter, there are no similar strong and relatively unblended narrow lines nearby to use for correction of the slight differences in the spectra from night to night.  We adopted an absolute flux calibration of $F(5007)=7.8\times10^{-13}$\,erg\,s$^{-1}$\,cm$^{-2}$, which is based on the values determined for NGC\,3227 over the last decade through a wide slit on photometric nights, as reported by \citet{derosa18} and \citet{brotherton20}.  

\begin{figure}
    \epsscale{1.2}
    \plotone{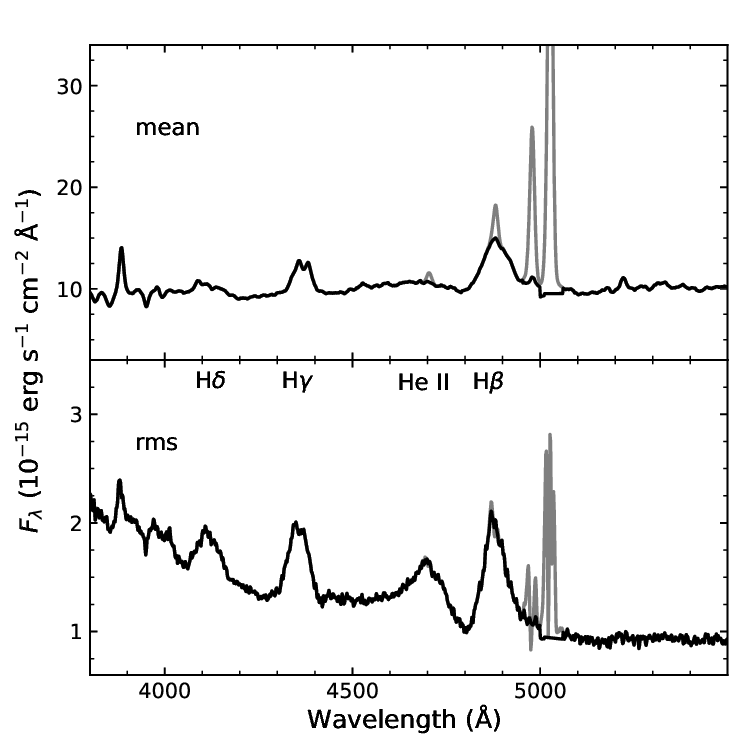}
    \caption{
    Mean (top) and rms (bottom) spectra of NGC\,3227 based on the observed-frame spectra obtained throughout the monitoring program.  The strong broad emission lines are labeled for ease of identification.
    }
    \label{fig:meanrms}
\end{figure}

Figure~\ref{fig:meanrms} displays the mean and root-mean-square (rms) of all the final calibrated and scaled spectra.  The rms spectrum highlights the variable spectral components, which include the AGN continuum and broad emission lines, while non-varying components like the narrow lines are eliminated.

\begin{deluxetable*}{lccccccc}
\renewcommand{\arraystretch}{1.2}
\tablecolumns{8}
\tablewidth{0pt}
\tablecaption{Emission-Line Time Lags and Widths}
\tablehead{
\colhead{} &
\colhead{} &
\colhead{} &
\colhead{} &
\multicolumn{2}{c}{mean} &
\multicolumn{2}{c}{rms}\\
\colhead{Line} &
\colhead{$\tau_{\rm cent}$} &
\colhead{$\tau_{\rm peak}$} &
\colhead{$\tau_{\rm jav}$} &
\colhead{FWHM} &
\colhead{$\sigma_{\rm line}$} &
\colhead{FWHM} &
\colhead{$\sigma_{\rm line}$}\\
\colhead{} &
\colhead{(days)} &
\colhead{(days)} &
\colhead{(days)} &
\colhead{(km s$^{-1}$)} &
\colhead{(km s$^{-1}$)} &
\colhead{(km s$^{-1}$)} &
\colhead{(km s$^{-1}$)}
}
\startdata
H$\beta$    & $4.03^{+0.85}_{-0.94}$ & $2.50^{+2.25}_{-1.25}$ & $2.671^{+0.126}_{-1.795}$ & $5070.3 \pm 69.1$  & $1943.1	\pm 12.5$  & $3710.0 \pm 186.4$ & $1682.3 \pm 39.2$  \\
H$\gamma$   & $2.73^{+0.61}_{-0.72}$ & $1.50^{+1.00}_{-0.25}$ & $1.156^{+0.237}_{-0.087}$ & $3860.7 \pm 62.5$  & $1548.2	\pm 21.5$  & $4313.3 \pm 179.0$ & $1566.0 \pm 60.5$   \\
H$\delta$   & $3.16^{+0.93}_{-1.00}$ & $1.50^{+1.50}_{-0.75}$ & $1.336^{+2.387}_{-0.165}$ & $4254.5 \pm 743.2$ & $1763.4	\pm 33.7$  & $4612.8 \pm 405.0$ & $1869.0 \pm 77.9$  \\
\ion{He}{2} & $0.94^{+1.07}_{-0.81}$ & $0.25^{+0.50}_{-0.25}$  & $0.003^{+0.685}_{-0.062}$ & \nodata  & \nodata  &     $6586.2	\pm 414.0$ & $2615.6 \pm 72.7$  
\label{tab:lagwidth}
\enddata

\tablecomments{Reported time lags are in the observer's frame, while, per convention, line widths are measured in the rest frame.}
\end{deluxetable*}

\begin{deluxetable*}{lccccccc}
\tablecolumns{8}
\tablewidth{0pt}
\tablecaption{Continuum and Emission-Line Light Curves}
\tablehead{
\multicolumn{2}{c}{Continuum} &
\colhead{} &
\multicolumn{5}{c}{Emission Lines} \\
\cline{1-2}
\cline{4-8}
\colhead{HJD} &
\colhead{$V$} &
\colhead{} &
\colhead{HJD} &
\colhead{H$\beta$} &
\colhead{H$\gamma$} &
\colhead{H$\delta$} &
\colhead{\ion{He}{2}} 
}
\startdata
59914.6126	& $7.566 \pm 0.111$ && 	59978.1643	& $5.747 \pm 0.105$ & $2.408 \pm 0.093$ & $0.725 \pm 0.101$ & $2.474 \pm 0.120$ \\
59915.2338	& $7.415 \pm 0.164$ && 	59980.2065	& $5.488 \pm 0.068$ & $2.335 \pm 0.061$ & $0.783 \pm 0.066$ & $2.345 \pm 0.078$ \\
59916.2424	& $6.597 \pm 0.149$ && 	59981.2027	& $5.752 \pm 0.063$ & $2.464 \pm 0.056$ & $0.676 \pm 0.062$ & $2.771 \pm 0.073$ \\
59917.0131	& $7.005 \pm 0.143$ && 	59984.1662	& $5.639 \pm 0.078$ & $2.365 \pm 0.070$ & $0.730 \pm 0.077$ & $2.508 \pm 0.090$ \\
59918.0019	& $7.340 \pm 0.148$ && 	59985.1536	& $5.766 \pm 0.058$ & $2.380 \pm 0.051$ & $0.752 \pm 0.056$ & $2.441 \pm 0.066$ 
\label{tab:lc}
\enddata 

\tablecomments{Heliocentric Julian dates are provided as HJD$-2400000$ (days).  $V-$band flux densities have units of $10^{-15}$\,erg\,s$^{-1}$\,cm$^{-2}$\,\AA$^{-1}$ while emission-line fluxes have units of $10^{-13}$\,erg\,s$^{-1}$\,cm$^{-2}$.  Table~\ref{tab:lc} is published in its entirety in the machine-readable format. A portion is shown here for guidance regarding its form and content.}
\end{deluxetable*}

\begin{deluxetable*}{lccccccc}
\tablecolumns{8}
\tablewidth{0pt}
\tablecaption{Light Curve Statistics}
\tablehead{
\colhead{Time Series} &
\colhead{$N$} &
\colhead{$\langle \Delta T \rangle$ (days)} &
\colhead{$\Delta T_{\rm med}$ (days)} &
\colhead{$\langle F \rangle$} &
\colhead{$\langle \sigma_F / F \rangle$} &
\colhead{$F_{\rm var}$} &
\colhead{$R_{\rm max}$}\\
\colhead{(1)} &
\colhead{(2)} &
\colhead{(3)} &
\colhead{(4)} &
\colhead{(5)} &
\colhead{(6)} &
\colhead{(7)} &
\colhead{(8)} 
}
\startdata
V          & 203 & $0.9 \pm 0.5$ & 0.75 & $6.24 \pm 0.61$ & 0.018 & 0.096 & $1.570 \pm 0.031$ \\
H$\beta$    & 51 & $2.4 \pm 2.8$ & 1.25 & $5.05 \pm 0.79$ & 0.013 & 0.157 & $1.735 \pm 0.033$ \\
H$\gamma$   & 51 & $2.4 \pm 2.8$ & 1.25 & $1.97 \pm 0.49$ & 0.030 & 0.249 & $2.173 \pm 0.086$ \\
H$\delta$   & 51 & $2.4 \pm 2.8$ & 1.25 & $0.57 \pm 0.30$ & 0.137 & 0.513 & $9.315 \pm 3.895$ \\
\ion{He}{2} & 51 & $2.4 \pm 2.8$ & 1.25 & $2.02 \pm 0.67$ & 0.040 & 0.334 & $4.294 \pm 0.301$ 
\label{tab:lcstats}
\enddata 

\tablecomments{$V-$band flux densities have units of $10^{-15}$\,erg\,s$^{-1}$\,cm$^{-2}$\,\AA$^{-1}$ while emission-line fluxes have units of $10^{-13}$\,erg\,s$^{-1}$\,cm$^{-2}$.}
\end{deluxetable*}

\vspace{-0.8in}

\section{Analysis} \label{sec:analysis}

\subsection{Line Width Measurements}

For each of the broad emission lines visible in the spectra of NGC\,3227, we report the line widths in both the mean spectrum and the rms spectrum.  Line widths were determined by setting a local linear continuum under each line and then measuring the distribution of flux directly from the data above the continuum within a specified wavelength window.  Line widths were quantified in two ways: the full width at half-maximum flux (FWHM) and the velocity dispersion of the line profile ($\sigma_{\rm line}$).  The uncertainties in the line widths were determined by selecting a random subset of spectra, building a new mean and rms spectrum from this subset, recording the FWHM and $\sigma_{\rm line}$, and repeating the process 1000 times to build up distributions of the measurements.  In Table~\ref{tab:lagwidth}, we report the median and inner 68\% confidence interval as the line width and its associated uncertainties, for both FWHM and $\sigma_{\rm line}$ as measured in the mean spectrum and the rms spectrum.  

We note that all reported line widths have been corrected for instrumental resolution by assuming that $\Delta \lambda_{\rm obs}^2 \approx \Delta \lambda_{\rm true}^2 + \Delta \lambda_{\rm disp}^2$ \citep{peterson04}.  The width of the [\ion{O}{3}] $\lambda 5007$\,\AA\ emission line was reported to be ${\rm FWHM}=485$\,km\,s$^{-1}$ when measured through a narrow slit at high spectral resolution by \citet{whittle92}, and we take this measurement to represent $\Delta \lambda_{\rm true}$.  The measured width of [\ion{O}{3}] $\lambda 5007$\,\AA\ in our spectra ($\Delta \lambda_{\rm obs}$) is ${\rm FWHM}=16.58$\,\AA.  We thus adopted a resolution correction of $\Delta \lambda_{\rm disp}=14.47$\,\AA\ or 866\,km\,s$^{-1}$.

\begin{figure}
    \epsscale{1.2}
    \plotone{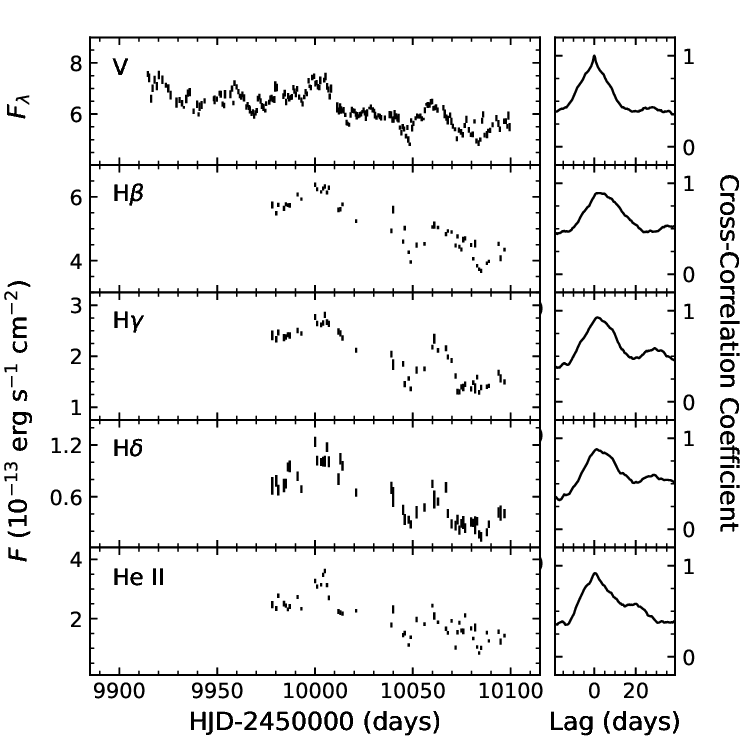}
    \caption{
    Light curves (left) for the continuum and broad emission lines throughout the course of the monitoring program.   Cross-correlation functions (right) are for each light curve relative to the continuum, and in the case of the continuum is the auto-correlation function.
    }
    \label{fig:lcs}
\end{figure}

\subsection{Emission-Line Light Curves and Time Delays} \label{sec:delays}

Light curves of the broad emission lines H$\beta$, \ion{He}{2}, H$\gamma$, and H$\delta$ were generated by fitting a local linear continuum under each line and integrating the flux above this continuum for each calibrated and scaled spectrum.  While simplistic, this method accounts for all of the flux in a line without having to rely on the discretion of the user in choosing an appropriate set of model parameters to properly fit asymmetric or complicated line shapes.  The non-varying narrow component of each line is included as a simple flux offset.  Table~\ref{tab:lc} tabulates the light curves. Table~\ref{tab:lcstats} lists basic information and variability statistics about each light curve; for each spectral feature listed in column (1), we give the number of measurements in column (2), and the average and median temporal sampling, respectively, in columns (3) and (4). Column (5) lists the mean flux and standard deviation, while column (6) lists the mean fractional error. Column (7) gives the noise-corrected fractional variation, which is computed as
\begin{equation}
    F_{\rm var} = \frac{\sqrt{\sigma^2 - \delta^2}}{\langle F \rangle}
\end{equation}
\noindent where $\sigma^2$ is the variance of the fluxes, $\delta^2$ is their mean-square uncertainty, and $\langle F \rangle$ is the mean flux. Column (8) lists the ratio of
the maximum to the minimum flux.

The typical time delay between each emission-line light curve and the continuum light curve was then assessed.  Throughout this work, we adopt the final $V-$band light curve from photometry as the continuum light curve because it has several advantages over the continuum as measured from the spectra.  In particular, the use of broad-band photometry, even on a small telescope, allows a high S/N to be reached in a short amount of observing time.  The wide field of view of the detectors available on small telescopes provide ample field stars that are observed simultaneously with the science target, providing a more accurate calibration than can generally be achieved for ground-based long-slit spectroscopy.  The abundance of small optical telescopes around the globe allows a significantly higher temporal sampling to be achieved, and the image subtraction methods employed in our analysis produce a high-fidelity light curve without the damping effect of nonvariable flux components.  Finally, the $V$ band was chosen because of its widespread availability and because it includes a mostly line-free portion of the spectrum for local AGNs like NGC\,3227.

Time delays were measured using the interpolated cross-correlation function (ICCF) method of \citet{gaskell86,gaskell87} with the modifications of \citet{white94}.  The cross-correlation function is determined twice with the ICCF method, first by interpolating one light curve and then the other.  The two resultant CCFs are averaged together to provide the final CCF, and we display these in Figure~\ref{fig:lcs} along with the continuum and emission-line light curves. 

The peak of the CCF ($r_{\rm max}$) occurs at the time shift for which the two light curves are most highly correlated.  This time delay is often reported as $\tau_{\rm peak}$.  A slight variation is to report $\tau_{\rm cent}$, which is the time delay associated with the centroid of the points around $\tau_{\rm peak}$ above some threshold (typically $0.8r_{\rm max}$).  We prefer $\tau_{\rm cent}$ because it is less susceptible to noise than is $\tau_{\rm peak}$ (cf.\ \citealt{peterson04}).

The uncertainties on $\tau_{\rm peak}$ and $\tau_{\rm cent}$ were derived following the flux randomization and random subset sampling (FR/RSS) method of \citet{peterson98b,peterson04}.  The FR/RSS method accounts for uncertainties in the time delay  arising from noise in the flux measurements as well as the inclusion (or exclusion) of each individual data point.  A large number of FR/RSS realizations ($N=1000$) build up distributions of $\tau_{\rm peak}$ and $\tau_{\rm cent}$ measurements.  From these distributions, we report the medians and inner 68\% confidence intervals as the time delay values and their associated uncertainties.  Time delays in the observed frame are tabulated in Table~\ref{tab:lagwidth}.  

We also measured the emission-line time delays using {\tt JAVELIN} \citep{zu11}.  {\tt JAVELIN} assumes a damped random walk for the behavior of the driving continuum light curve, and optimizes the parameters for a top-hat transfer function to predict the responding emission-line light curve(s).  Even though {\tt JAVELIN} is capable of fitting multiple emission-line light curves at the same time, we were unsuccessful in fitting them all simultaneously, so we instead fit each one separately.  The best-fit time delays are listed in Table~\ref{tab:lagwidth}, and while they agree well with $\tau_{\rm peak}$, they generally do not agree with $\tau_{\rm cent}$.  Differences  between $\tau_{\rm peak}$ and $\tau_{\rm cent}$ are caused by the shape of the CCF, which is in turn related to the transfer function.  We speculate that in this case, {\tt JAVELIN}'s assumption of a top hat may not be a good match to the actual transfer function of NGC\,3227.

\begin{figure}
    \epsscale{1.15}
    \plotone{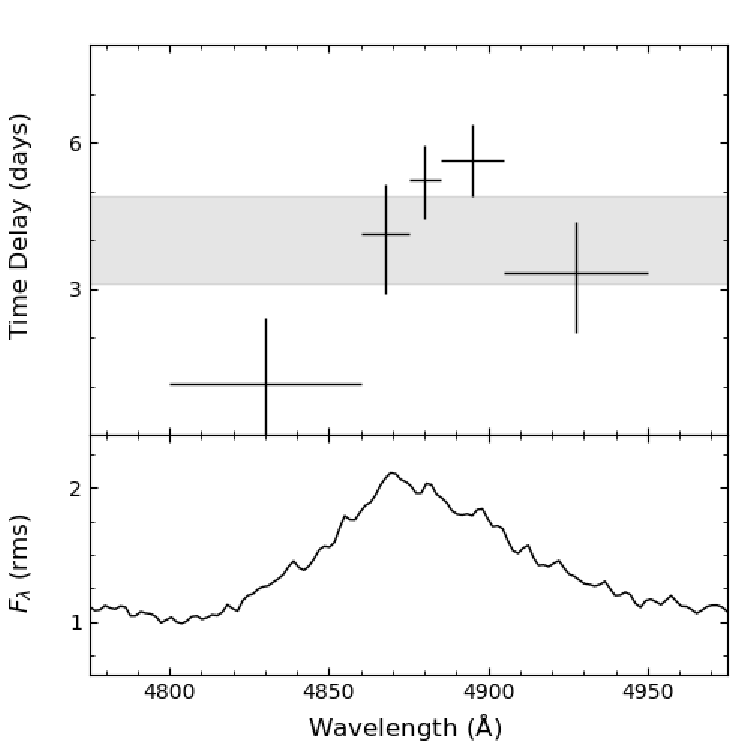}
    \caption{
    Velocity-resolved time delays (top) across the broad H$\beta$ emission-line profile (bottom).
    }
    \label{fig:velres}
\end{figure}

Using the same methods as described above, we also divided the H$\beta$ broad emission line into 5 separate bins, with similar flux in all bins, and generated a light curve for each one to assess the H$\beta$ gas response as a function of line-of-sight velocity.  The time delay between the continuum and the light curve associated with each bin was then determined through cross correlation following the same methods outlined above.  As we show in Figure~\ref{fig:velres}, there are clear differences in the time delay associated with each bin across the broad emission-line profile, with the longest time delays near the line center and the shortest time delay associated with the blue wing.  This velocity-resolved behavior is qualitatively similar to what has been observed in previous velocity-resolved analyses of NGC\,3227 \citep{denney10,derosa18,brotherton20}.

\begin{deluxetable*}{LlC}
\renewcommand{\arraystretch}{1.1}
\tablecolumns{3}
\tablecaption{Broad-line region model parameter values}
\tablehead{
\colhead{Parameter} & 
\colhead{Brief Description} & 
\colhead{H$\beta$} 
}
\startdata
\log_{10} (M/M_{\odot})         & Black hole mass                       & 7.09^{+0.35}_{-0.34}       \\
r_{\rm mean}\ \rm(light~days)   & Mean radius of line emission          & 5.50^{+1.64}_{-1.33}      \\ 
r_{\rm median}\ \rm(light~days)  & Median radius of line emission        & 5.03^{+1.79}_{-1.44}      \\
r_{\rm min}\ \rm(light~days)     & Minimum radius of line emission       & 2.54^{+3.20}_{-1.65}       \\
\sigma_{r}\ \rm(light~days)      & Radial extent of line emission        & 2.33^{+5.23}_{-1.10}    \\
\tau_{\rm mean}\ \rm(days)       & Mean time delay                       & 4.93^{+0.64}_{-0.67}       \\
\tau_{\rm median}\ \rm(days)     & Median time delay                     & 3.81^{+0.56}_{-0.52}       \\
\theta_{o}\ \rm(degrees)        & Opening angle                         & 64.7^{+18.3}_{-11.8}         \\
\theta_{i}\ \rm(degrees)        & Inclination angle                     & 33.2^{+13.5}_{-9.1}          \\
\beta                           & Shape parameter of radial distribution & 0.84^{+0.64}_{-1.14}      \\
\gamma                          & Disk face concentration parameter     & 3.95^{+0.75}_{-1.14}       \\
\xi                             & Transparency of the mid-plane         & 0.49^{+0.38}_{-0.30}       \\
\kappa                          & Cosine illumination function parameter & -0.28^{+0.19}_{-0.15}      \\
f_{\rm ellip}                   & Fraction of elliptical orbits         & 0.71^{+0.14}_{-0.37}       \\
f_{\rm flow}                    & Inflow vs.\ outflow                   & 0.45^{+0.38}_{-0.30}       \\
\theta_{e}\ \rm(degrees)         & Ellipse angle                         & 57.5^{+12.1}_{-26.8}       \\
\sigma_{\rm turb}               & Turbulence                            & 0.006^{+0.018}_{-0.004}    \\
r_{\rm out}\ \rm(light~days)     & Outer radius of line emission (fixed parameter) & 78              \\
T                               & Temperature or likelihood softening   & 65                       
\label{tab:modelpars}
\enddata

\tablecomments{Tabulated values are the median and 68\% confidence intervals.}
\end{deluxetable*}

\begin{figure*}
    \epsscale{1.2}
    \plotone{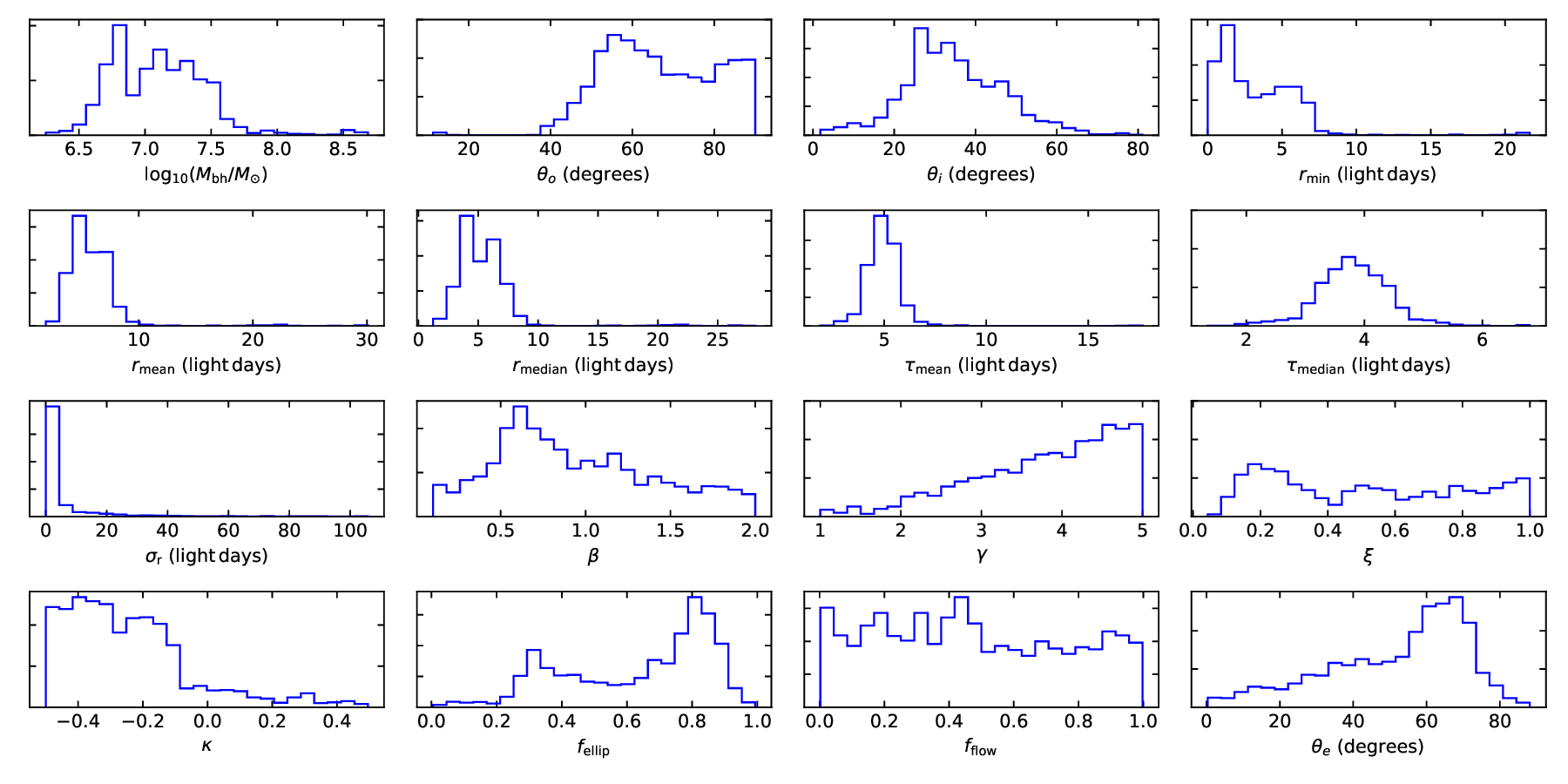}
    \caption{
    Histograms displaying the posterior distributions of the BLR model parameters for H$\beta$.
    }
    \label{fig:modelpars}
\end{figure*}

\vspace{-0.25in}

\subsection{{\tt CARAMEL} Modeling}

Using the full velocity-resolved response of an emission line to constrain the geometry and kinematics of the BLR is usually approached in two ways: through the inverse approach, in which deconvolution of the emission-line response produces a velocity-delay map (e.g., \citealt{horne04,skielboe15,anderson21}), or through forward modeling, in which a self-consistent set of models are used to predict the emission-line response and then compared to the observations (e.g., \citealt{pancoast14a,rosborough23}).  The first approach  has the advantage of relying on a relatively small set of assumptions and has the ability to make use of all the details in the data, but short baselines, irregular sampling, and noisy data complicate the deconvolution process, and the results are difficult to interpret without extensive modeling.  The second approach is relatively easy to interpret, but is fundamentally  limited by a larger set of core assumptions and the level of flexibility and completeness in the models. 

{\tt CARAMEL} is a phenomenological modeling code that was developed to explore the BLR using reverberation data sets.  It is described in detail by \citet{pancoast14a} and we have previously used it in the study of other nearby AGNs \citep{bentz21c,bentz22,bentz23}.  By exploring the parameter space of geometric and kinematic models that may represent the BLR, {\tt CARAMEL} is able to provide a direct and primary constraint on the black hole mass (without resorting to the use of a scale factor) and can also provide insight into various characteristics of interest, such as the inclination of the BLR to our line of sight.  In the few cases where independent measurements are available, they have been to found to be in general agreement with {\tt CARAMEL} results, such as the mass of the black hole and the BLR inclination in NGC\,4151 \citep{bentz22}, and the general structural parameters of the BLR in NGC\,3783 \citep{gravity21,bentz21c}.

We followed the approach described in our previous work \citep{bentz21c,bentz22,bentz23} and first carried out a  decomposition of each spectrum using the {\tt ULySS} package \citep{koleva11}.  We modeled all of the continuum and emission-line features in the spectra and then subtracted the continuum components and any emission lines close in wavelength to H$\beta$, leaving an isolated H$\beta$ feature corresponding to each observational epoch.  These isolated observed H$\beta$ profiles were then fed into {\tt CARAMEL} for comparison with the line profiles created from each model, along with the driving continuum light curve.

\begin{figure}
    \epsscale{1.15}
    \plotone{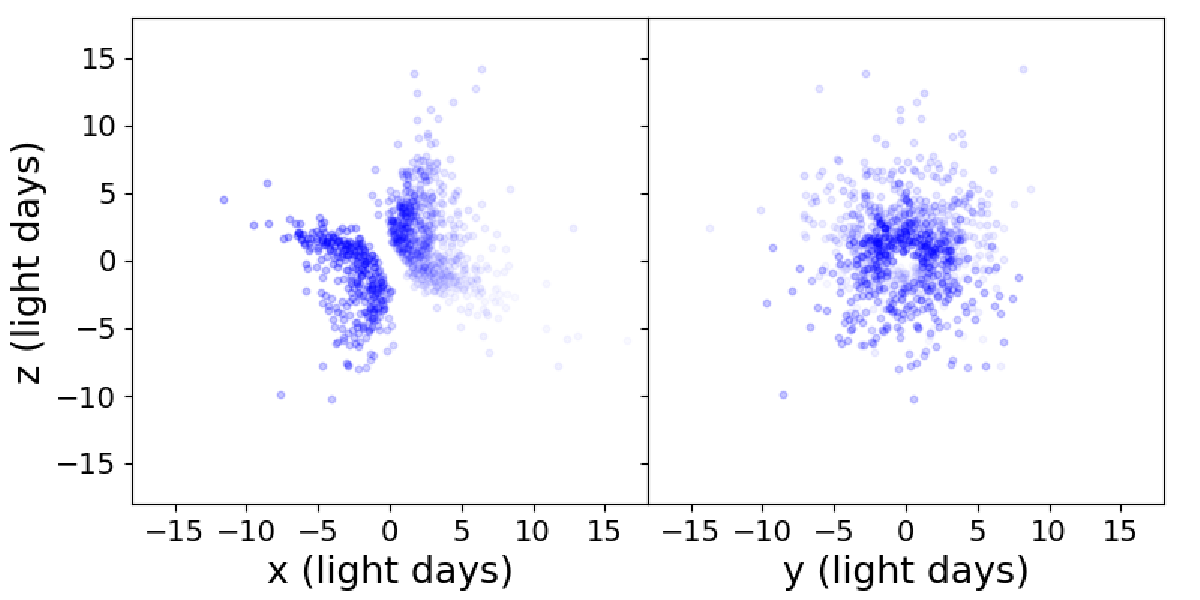}
    \caption{
    Representative geometric model of the H$\beta$ emission in the BLR of NGC\,3227.  The left panel is oriented edge on, with an Earth-based observer at large  $+x$ values, and the right panel shows the Earth-based observer’s view.  The transparency of each point is related to the strength of the response of the gas, with more opaque points having a stronger response to continuum fluctuations.
    }
    \label{fig:blrclouds}
\end{figure}

In Figure~\ref{fig:modelpars} we display the posterior probability distribution function for the model parameters preferred by {\tt CARAMEL}, and we tabulate the median and 68\% confidence interval for each parameter in Table~\ref{tab:modelpars}.  The median values of these parameters indicate that the H$\beta$-emitting region of the BLR may be represented by the surface or ``skin'' of a biconical or flared disk structure with its axis of symmetry inclined to our line of sight at an angle of $\theta_i=33.2^{+13.5}_{-9.1}$\,degrees, with preferential radiation back towards the central source ($\kappa = -0.28^{+0.19}_{-0.15}$), and with gas motions that are dominated by rotation ($f_{\rm ellip}=0.71^{+0.14}_{-0.37}$).  We include an illustrative example of this geometry in Figure~\ref{fig:blrclouds}, and note that strong inward anisotropy of hydrogen emission from the BLR is expected due to the high densities and correspondingly large line optical depths in the gas \citep{ferland92}.

\section{Discussion}

\subsection{Radius vs.\ Luminosity}

Previous measurements of the H$\beta$ time delay in NGC\,3227 range from $\sim 2$\,days \citep{derosa18} to $\sim 7$\,days \citep{brotherton20}.  The time delay of H$\beta$ is expected to change as a function of the central AGN luminosity.  The relationship for these two quantities follows a power-law shape with a slope of $\approx0.5$ for a sample of local Seyferts (the \rl\ relationship; \citealt{bentz06a,bentz09b,bentz13}), and each individual AGN is expected to have its own \rl\ relationship as its luminosity varies (e.g., \citealt{bentz07,kilerci15}), so we would not expect to always measure the same time delay for a specific emission line in a single AGN. 

Nevertheless, the time delay that we find for H$\beta$ of $\tau_{\rm cent}=4.03^{+0.85}_{-0.94}$\,days is nearly the same as the time delay reported by \citet{denney10}, $\tau_{\rm cent}=3.76^{+0.76}_{-0.82}$\,days.  We find a mean continuum flux of $f_{\rm cont}=(9.56 \pm 0.04)\times 10^{-15}$\,erg\,s$^{-1}$\,cm$^{-2}$\,\AA$^{-1}$ at $5100 \times (1+z)$\,\AA.  After correcting for the host-galaxy starlight contribution to the continuum, $f_{\rm gal}= (5.1 \pm 0.5) \times 10^{-15}$\,erg\,s$^{-1}$\,cm$^{-2}$\,\AA$^{-1}$, derived from high-resolution Hubble Space Telescope imaging following the methods of \citet{bentz09b,bentz13}, we find an AGN luminosity of $\log (\lambda L_{\rm 5100}/L_{\odot}) = 42.21 \pm 0.11$.  This is very similar to the luminosity reported by \citet{denney10}, after correction for the distance of $D=23.7\pm2.6$\,Mpc assumed here, of  $\log (\lambda L_{\rm 5100}/L_{\odot}) = 42.24 \pm 0.12$.  Interestingly, both of these measurements lie exactly on the best-fit \rl\ relationship of \citet{bentz13}.

On the other hand, the two H$\beta$ time delay measurements reported by \citet{derosa18} lie below the \rl\ relationship, while the H$\beta$ time delay reported by \citet{brotherton20} lies above it, even after ensuring uniformity in the assumed distances, Galactic extinction, and method of starlight correction when deriving the AGN luminosity. The longer time delay reported by \citet{brotherton20}, when compared to the measurements presented here and by \citet{denney10}, seems to tentatively agree with the expectation of a steeper slope for the \rl\ relationship of a single AGN when the optical luminosity is used as a proxy for the ionizing luminosity \citep{kilerci15}.  However, the AGN luminosity was quite similar during the monitoring program of \citet{brotherton20} and the programs described by \citet{derosa18}, yet the time delays that were derived are quite different.  We note that the amplitude of variability relative to the noise was significantly lower in the light curves presented by \citet{derosa18}.  If the measurements of $\tau_{\rm cent}$ and their larger uncertainties are preferred over those of $\tau_{\rm jav}$, then the measurements of \citet{derosa18} are within the typical scatter around the \rl\ relationship, and they agree with the time delay measurement reported by \citet{brotherton20} at the $3\sigma$ level.

\subsection{Black Hole Mass} \label{sec:mbh}

Because of its proximity, NGC\,3227 is one of the few Seyfert galaxies that has  previous dynamical mass measurements.  We note that dynamical masses are linearly dependent on the adopted distance to the galaxy, so any comparison of masses from different techniques must keep this detail in mind.  NGC\,3227 is near enough that its redshift of $z=0.00377$ cannot be used to accurately predict its distance.  Instead, throughout this work we have adopted the SBF-based distance to NGC\,3226 as the distance to NGC\,3227 (given their clear ongoing interaction).  The SBF distance of $D=23.7\pm2.6$\,Mpc is based on the analysis of \citet{tonry01}, with slight adjustments by \citet{blakeslee01}.  In addition, a careful assessment of the predicted distance based on the \citet{tully77} relationship by \citet{robinson21} finds $D=24.3\pm4.9$\,Mpc for NGC\,3227, which is in good agreement.

Previous mass constraints from stellar dynamical modeling \citep{davies06} and  from gas dynamical modeling \citep{hicks08} used different distances from our adopted value and from each other, both relying on various measurements of the redshift instead.  Adjusting the dynamical masses to account for their assumed distances relative to our adopted distance gives a stellar dynamical mass of $M_{\rm BH} = (1.9\pm0.9)\times10^7$\,$M_{\odot}$ and a gas dynamical mass of $M_{\rm BH} = (3.1^{+1.5}_{-0.6})\times10^7$\,$M_{\odot}$. While these values agree with each other, they have generally not agreed with reverberation-based masses, being a factor of $4-5$ larger \citep{denney10,derosa18}.

For most reverberation analyses, black hole mass is calculated from the time delay ($\tau$) and line width ($V$) as:
\begin{equation}
    M_{\rm BH} = f \frac{c\tau V^2}{G}
\end{equation}
\noindent where $c$ is the speed of light, and $G$ is the gravitational constant.  An extensive analysis of reverberation datasets and analysis procedures by \citet{peterson04} provided a list of best practices, particularly recommending the use of $\tau_{\rm cent}$ for the time delay and $\sigma_{\rm line}(rms)$ for the line width.  The scale factor $f$ encodes all the geometric and kinematic details of the BLR, including the inclination angle to our line of sight.  Since these details are generally unknown for individual AGNs, a population-average scale factor, $\langle f \rangle$, is usually adopted so that the \msigma\ relationship for AGNs with masses from reverberation is brought into general agreement with the relationship for mostly quiescent galaxies with black hole masses from dynamical modeling (e.g., \citealt{onken04}).  For the combination of $\tau_{\rm cent}$  and $\sigma_{\rm line}(rms)$, the value of $\langle f \rangle$ has been found to range from 2.8 \citep{graham11} to 5.5 \citep{onken04}, depending on the exact details and the sample that is included.  We generally prefer to adopt $\langle f \rangle=4.8$ from \citet{batiste17b} because of their careful treatment of the effects of galaxy morphology, including bars, on the determination of $\sigma_{\star}$.  

While in practice, the adoption of $\langle f \rangle$ minimizes any bias in the full sample of reverberation masses, it also means that the reverberation mass for any particular AGN may be over/underestimated by a factor of a few.  If we assume that the variance in $f$ values between different AGNs is mainly due to their random inclinations to our line of sight (as suggested by the analysis of \citealt{williams18}), then $\langle f \rangle \approx 5$ implies a typical inclination of 27\degr.  For an AGN that is observed at a lower (more face-on) inclination, $\langle f \rangle \approx 5$ would significantly underestimate the correction needed to convert the line-of-sight velocity component to the true velocity when determining \mbh.  And in the case of NGC\,3227, a factor of $\sim 4$ discrepancy between the reverberation mass and the dynamical masses could be resolved if the AGN were found to be viewed at an angle of $\sim15$\degr.

Using the measurements for H$\beta$ listed in Table~\ref{tab:lagwidth} indicates $M_{\rm BH}=1.07^{+0.23}_{-0.25} \times 10^7$\,\msun.  This is somewhat larger than the masses that were determined by \citet{denney10} and \citet{derosa18}, but the 1$\sigma$ range formally agrees with the low end of the uncertainties on the much higher mass reported by \citealt{brotherton20}.  The time delay for H$\beta$ found in this work and the analysis of \citet{denney10} are nearly identical, however, somewhat perplexingly, while the FWHM measurements of H$\beta$ agree within the uncertainties, $\sigma_{\rm line}$ is $\sim 300$\,km\,s$^{-1}$ (22\%) larger in our measurements leading to a larger derived mass.  The rms line profile presented by \citet{denney10} shows a clear double-peaked shape without strong wings, while the rms line profile in our observations has strong wings and the same blue peak but without the corresponding red peak, potentially indicating a significant change in the properties of the reverberating gas.  

The H$\beta$ mass from our time delay and line width measurements is in good agreement with the mass from {\tt CARAMEL} modeling of $\log M/M_{\odot} = 7.09^{+0.35}_{-0.34}$ or $M_{\rm BH}=1.23^{+1.52}_{-0.67} \times 10^7$\,\msun, and in this case, the good agreement is likely because the inclination angles preferred by {\tt CARAMEL} ($\theta_i = 33.2^{+13.5}_{-9.1}$) are similar to the population average value implied by $\langle f \rangle \approx 5$.  Interestingly, this inclination angle agrees with the results of a reanalysis of the geometry of the spatially resolved narrow line region \citep{falcone23}, while an earlier analysis suggested a much smaller value of $\theta_i\approx15\degr$ \citep{fischer13}.

Additionally, while the black hole mass from this reverberation program agrees with the dynamical masses, there are still good reasons to revisit the dynamical constraints.  Gas dynamical modeling is known to be susceptible to biases when the gas exhibits noncircular motions, and this worry should be taken seriously in the presence of an AGN (e.g., \citealt{verdoes06,jeter19}).  While stellar dynamical modeling is usually held to be more accurate, the stellar dynamical modeling analysis of \citet{davies06} relied on fairly shallow but high spatial resolution observations of the nucleus of NGC\,3227, and the $S/N$ of the observations only allowed $V$ and $\sigma$ to be constrained for the stellar kinematics.  Best practices in stellar dynamical modeling require that the kinematic constraints include not only $V$ and $\sigma$, but also additional higher order terms to reduce the degeneracy between velocity anisotropy of the stellar distribution and \mbh\ \citep{binney82}.  Because of this potential degeneracy, deeper observations that provide more detailed information regarding the nuclear stellar kinematics are needed to accurately assess the dynamical constraints on the black hole mass in NGC\,3227.

\subsection{BLR Structure and Kinematics} 

Modeling of the velocity-resolved H$\beta$ response in NGC\,3227 with {\tt CARAMEL} suggests that the BLR may be represented by the surface of a biconical structure or flared disk, oriented at $\sim 33\degr$ to our line of sight and with gas motions that are dominated by rotation around a black hole with $M_{\rm BH} \approx 1.2 \times 10^7$\,\msun. While there are independent measurements of the inclination and black hole mass with which we can compare the {\tt CARAMEL} constraints, as we have described above, there are few options for independent tests of the other BLR properties determined by {\tt CARAMEL}.

One promising new tool that is currently being developed is {\tt BELMAC} (Broad Emission-Line Mapping Code; \citealt{rosborough23}), an extension of  {\tt TORMAC} \citep{almeyda17,almeyda20} that has been adapted for application to the BLR.  {\tt BELMAC} simulates the velocity-resolved response of an emission line to an input driving light curve.  Using photoionization grids and a 3D ensemble of gas clouds, {\tt BELMAC} simulates the observed response of the gas clouds to continuum variations for various user-specified geometries, velocity fields, and cloud properties. 

While {\tt BELMAC} is not yet able to independently explore and optimize the models to determine a best fit to the observed behavior of an emission line (though that capability is currently under development), we were able to use it in its current form to carry out two important tests: (1) when given the results of the {\tt CARAMEL} models, could {\tt BELMAC} confirm that the models provided a good fit to the data?, and (2) what types of geometric and kinematic parameters allow {\tt BELMAC} to most closely replicate the observed integrated H$\beta$ light curve and mean and rms line profiles?

\begin{figure}
    \epsscale{1.15}
    \plotone{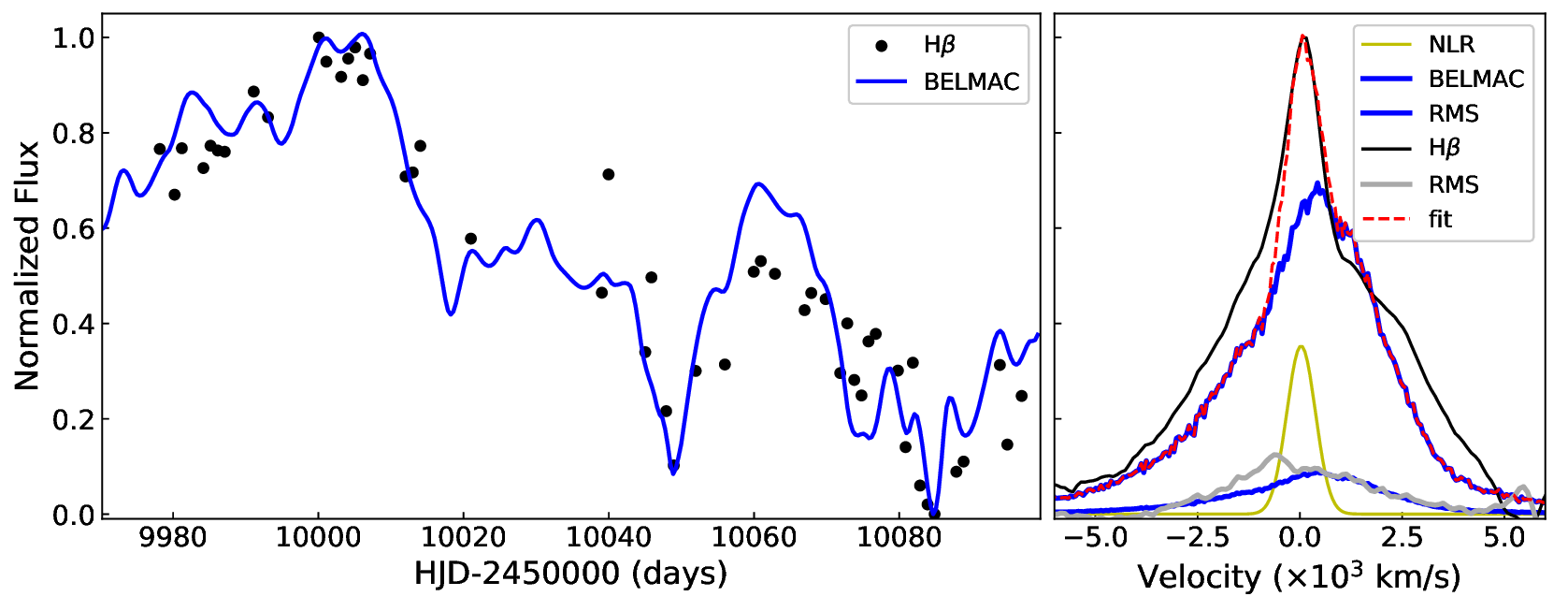}
    \caption{H$\beta$ and model light curves (left) as black points and blue curve, respectively. Mean and rms line profile in black and grey, respectively, and the model mean and rms line profiles in blue and light blue, respectively, for broad H$\beta$.  The narrow H$\beta$ line profile is in yellow and the overall fit is in red (right).  The light curve and line profile models were created using {\tt BELMAC} with parameters interpreted from Table~\ref{tab:modelpars}.   
    }
    \label{fig:cartobel}
\end{figure}

\begin{figure}
    \epsscale{1.15}
    \plotone{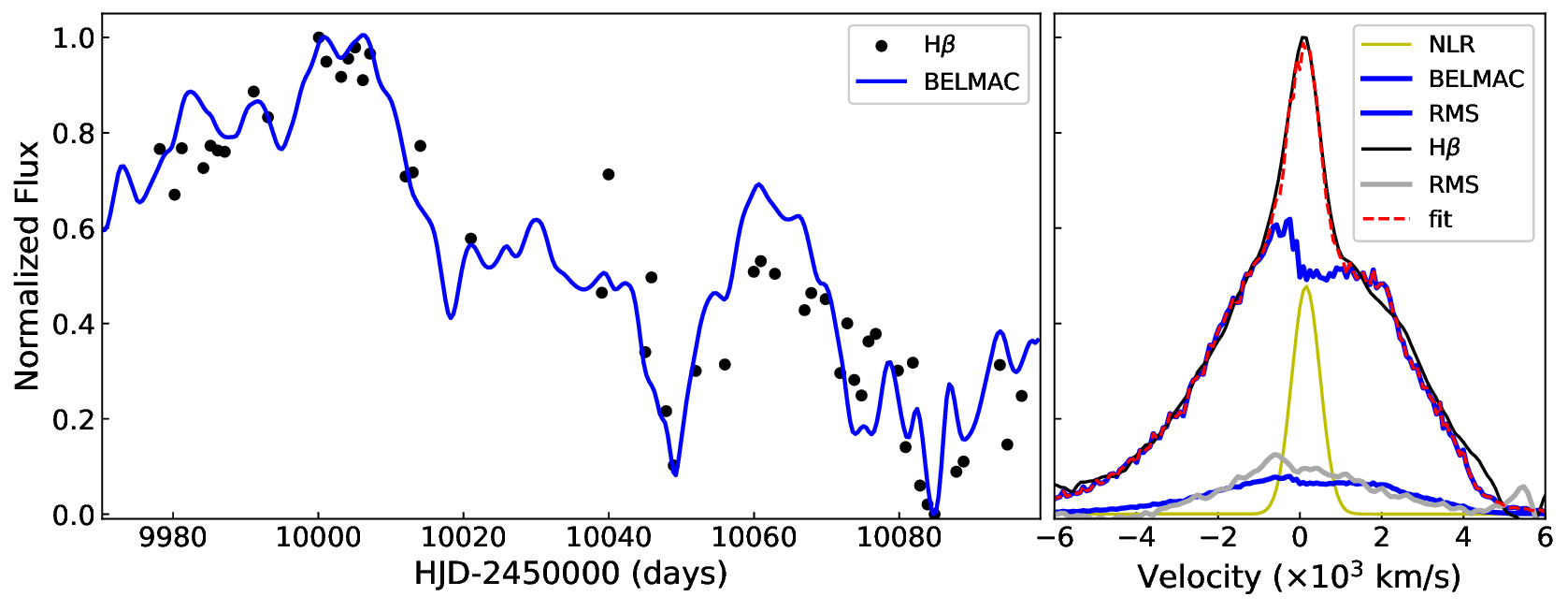}
    \caption{The same as Figure~\ref{fig:cartobel}, but with model parameters adjusted to best match the H$\beta$ light curve, line profile, and rms profile.   
    }
    \label{fig:belmac}
\end{figure}

We began by calculating the photoionization grids using CLOUDY v17.03 \citep{ferland17} assuming a bolometric luminosity for NGC\,3227 of  $\log L_{\rm bol}=43.16$\,erg\,s$^{-1}$ (where $L_{\rm bol}=9 \lambda L_{\rm 5100}$; \citealt{kaspi00,merritt22}) and an SED based on the average of 17 local, unobscured AGNs with Eddington ratios $<0.1$ from \citet{jin12}. The {\tt BELMAC} model results shown in Figure~\ref{fig:cartobel} were then created using the median values of $r_{min}$, $r_{mean}$, $\theta_i$, and \mbh\ listed in Table~\ref{tab:modelpars}.  Although $\gamma$ and $\xi$ are not {\tt BELMAC} parameters, they imply a biconical geometry, as shown in Figure~\ref{fig:blrclouds}, which is a {\tt BELMAC} geometry configuration.  To obtain a better match to the H$\beta$ data, $\theta_o$ was used as a guide rather than a fixed parameter.  Assuming a bicone, $\theta_o$ is approximately the separation angle between each cone and the midplane.  Therefore, the angular distribution of clouds ranged from $5-55\degr$, where $0\degr$ is the midplane, for each cone.  Similarly, the emission distribution described by {\tt CARAMEL's} $\beta$ and $\sigma_r$ parameters provide estimates for the cloud distribution, gas density distribution, and covering fraction.  In {\tt BELMAC}, the number density of clouds with radius and gas density of the clouds with radius are, respectively, $N(r)\propto r^p$ and $n_H(r)\propto r^{-s}$, where $p$ and $s$ are the free parameters.  The cross-sectional area of the clouds is primarily determined from the covering fraction.  To approximately match {\tt CARAMEL's} emission distribution, we adopted $p=-2$, $s=-1.45$, with a gas density at the outer radius of $\log n_H =9.3$\,cm$^{-3}$, and a covering fraction of 0.4.  Lastly, rotational and radial motions were included in the {\tt BELMAC} model, where the radial cloud motion is influenced by radiation pressure and gravity.  Therefore, using the {\tt CARAMEL} parameters listed in Table~\ref{tab:modelpars} to guide the above described {\tt BELMAC} parameters, the radial motion is entirely an inflow. In general, the model is able to reproduce the integrated light curve fairly well, and does a reasonable job matching the mean and rms line profiles, though there is some mismatch in the wings of the mean line profile and the blue peak of the rms profile.

In comparison, Figure~\ref{fig:belmac} shows the best {\tt BELMAC} model results to the data. Since there is currently no parameter optimization capability, the parameters were manually adjusted until a reasonable match to the data was found.  However, the same $r_{min}$, $r_{mean}$, and \mbh\ determined by {\tt CARAMEL} were used, as well as the same photoionization grids.  The best model found is a disk with a half-angular width $30\degr$ (analogous to $\theta_o$) and inclined to our line of sight at $35\degr$.  About $53\%$ of the clouds in the model have bound elliptical orbits and $47\%$ are unbound hyperbolic orbits, of which  $98\%$ are inflowing. The geometry is described by $p=-1.7$, $s=-1.3$, where the gas density at the outer radius is $\log n_H=9.2$\,cm$^{-3}$, and the covering fraction is 0.33.  Other than the overall shape of this {\tt BELMAC} BLR model, which is a thick disk rather than a bicone, and a preference for a stronger inflow component, the parameters are in general agreement with those inferred by {\tt CARAMEL}, lending additional credence to the derived inclination angle and black hole mass.  Furthermore, while some of the details appear to be slightly different between the preferred models from the two codes, we note that it is not straightforward to directly compare them since {\tt CARAMEL} models the emission itself while {\tt BELMAC} includes a full distribution of gas clouds, including those that may be shielded or only weakly emitting.  Nevertheless, both agree that the H$\beta$-emitting BLR is extended in radius and of similar predicted size, that the geometry is flattened rather than spherical, and that the kinematics are not dominated by outflows (as might be expected for a disk wind) but instead are dominated by rotation and/or inflow.

\section{Summary}

We have carried out a new optical reverberation mapping campaign focused on the nearby broad-lined Seyfert  NGC\,3227.  Using the LCO global network of telescopes, we employed broad-band photometry to track the continuum variations coupled with spectroscopy to track the responses of the broad emission lines.  Time delays relative to the continuum variations were found for the H$\beta$, \ion{He}{2}, H$\gamma$, and H$\delta$ broad emission lines.  We also detected velocity-resolved time delays across the broad H$\beta$ emission-line profile.  

Modeling of the velocity-resolved H$\beta$ response with {\tt CARAMEL} finds a black hole mass of $\log M/M_{\odot} = 7.09^{+0.35}_{-0.34}$ or $M_{\rm BH}=1.2^{+1.5}_{-0.7} \times 10^7$\,\msun, which agrees well with the mass determined from the H$\beta$ time delay and line width when atypical scaling factor is assumed, $M_{\rm BH}=1.1^{+0.2}_{-0.3} \times 10^7$\,\msun.  This good agreement is likely due to the intermediate inclination that is preferred by the models, $\theta_i\approx33\degr$, which is similar to the population-average inclination angle that is implied by typical RM scale factors of $\langle f \rangle=4-5$.  The models also suggest that the H$\beta$-emitting BLR may be represented by the inner surface of a biconical or flared disk structure, and that the gas motions are dominated by rotation.

Finally, using the photoionization-based modeling code {\tt BELMAC}, we confirmed that the {\tt CARAMEL} results provide a reasonable fit to the H$\beta$ light curve and mean and rms line profiles, though there is some mismatch in the line wings.  To correct for this mismatch, {\tt BELMAC} prefers a thick disk geometry, rather than a bicone, and roughly equal weight between rotation and inflow.  While they disagree about these details, both codes agree on several general properties, including that the H$\beta$-emitting BLR is extended in radius, flattened rather than spherical, and that the kinematics are dominated by rotation and/or inflow rather than outflow.  Future improvements to the {\tt BELMAC} code will allow a more thorough exploration of reasonable BLR parameter combinations for this AGN and others, and a more detailed comparison with the modeling results from {\tt CARAMEL}. 

\begin{acknowledgements}
MCB gratefully acknowledges support from the NSF through grant AST-2009230. MM was supported by National Science Foundation grant No.\ 2050829 to Georgia State University under the Research Experiences for Undergraduates program. SR was supported by NSF grant AST-2009508.  CAO was supported by the Australian Research Council (ARC) through Discovery Project DP190100252. MV was supported by the NSF through AST-2009122. TT gratefully acknowledges supprt by NSF through grant AST-1907208.

\end{acknowledgements}


\facility{LCOGT}

\software{LCO BANZAI Reduction Pipeline \citep{mccully18}, Galfit \citep{peng02,peng10}, JAVELIN \citep{zu11}, ULySS \citep{koleva11}, CARAMEL \citep{pancoast14a},
BELMAC \citep{rosborough23}, CLOUDY v17 \citep{ferland17} }


\end{document}